\documentclass[prl,amsmath,amssymb,reprintnumbers,showpacs, nofootinbib]{revtex4}
\usepackage{latexsym}
\usepackage{epsfig}
\usepackage{amssymb}
\usepackage{amsmath}
\usepackage{color}
\usepackage{subfigure}
\usepackage{comment}
 \usepackage[scaled=1]{helvet}

\newcommand{\be}{\begin{eqnarray}}
\newcommand{\ee}{\end{eqnarray}}

\begin{document}

\preprint{}
\title{Nonadditive Entropies Yield Probability Distributions with Biases not Warranted by the Data}

\author{Steve Press\'e$^{1}$ $^{*}$, Kingshuk Ghosh$^{2}$, Julian Lee$^{3}$, Ken A. Dill$^{4}$}
\date{\today}
\affiliation{ 
$^{*}$ Corresponding author.\\
$^1$ Department of Physics, Indiana Univ. -Purdue Univ., Indianapolis, Indianapolis, IN 46202. \\
$^2$ Department of Physics and Astronomy, University of Denver, CO 80208.\\
$^3$ Department of Bioinformatics and Life Science, Soongsil University, Seoul 156-743, Korea.\\
$^4$ Laufer Center for Physical and Quantitative Biology, Stony
Brook University, NY 11794. }

\begin{abstract}
Different quantities that go by the name of entropy are used in variational principles to infer probability distributions from limited data.  Shore and Johnson showed that maximizing the Boltzmann-Gibbs form of the entropy ensures that probability distributions inferred satisfy the multiplication rule of probability for independent events in the absence of data coupling such events.  Other types of entropies that violate the Shore and Johnson axioms, including nonadditive entropies such as the Tsallis entropy, violate this basic consistency requirement. Here we use the axiomatic framework of Shore and Johnson to show how such nonadditive entropy functions generate biases in probability distributions that are not warranted by the underlying data. \\
{\bf Please cite: S. Press\'e, K. Ghosh, J. Lee, K.A. Dill ``Nonadditive Entropies Yield Probability Distributions with Biases not Warranted by the Data", Phys. Rev. Lett., 111, 180604 (2013)}.
\end{abstract}

\maketitle

A problem of broad interest across the sciences is to infer the mathematical form of a probability distribution given limited data \cite{rmp}.  
For instance, we may be given limited information on an equilibrium system --say its average energy-- 
from which we must predict the mathematical form of the full energy probability distribution. 
In this classic example, the distribution used in statistical mechanics is the exponential Boltzmann distribution. 

In many cases --including the case above-- limited data is consistent with many possible models for a probability distribution.  
How should we select the ``best'' model that fits the data?  
By model, we mean a set of probablities, $\{ p_{k}\}$, for the outcomes $k$ of an experiment.  One way to select the model is to eliminate candidate models by supplementing the data with additional assumptions ~\cite{live, skilling, skilling91, xie2002, skillingaxiom}.  A common additional assumption is based on choosing the model that has the largest entropy.  This is basis for the variational principle called Maximum Entropy (MaxEnt)~\cite{shannon}. 

The MaxEnt approach has its historical roots in the work of Boltzmann in equilibrium statistical physics.  Later, Shannon and Jaynes showed that picking the model with the largest entropy was analogous to maximizing the uncertainty, $H$, of the model ~\cite{shannon, jayneslogic, jaynes57, jaynes57part2}.  In particular, Jaynes drew the connection between statistical mechanics and Shannon's work on information theory by showing that since Shannon's $H = - \sum p_k \log p_k$ coincides with the Boltzmann-Gibbs (BG) entropy,  statistical mechanics could be treated as an inference problem ~\cite{jaynes57, jaynes57part2}. 
Shannon and Jaynes and others justified the specific mathematical form, $H = - \sum p_k \log p_k$ on the basis of abstract properties of $H$ itself, such as satisfying a composition property ~\cite{shannon}. Others justified the form for $H$ using the Khinchin axioms ~\cite{khinchin, hanel} or the Fadeev postulates~\cite{renyi}, for example.  

By contrast to these methods for deriving $H$ on the basis of $H$'s properties, Shore and Johnson (SJ)~\cite{shore} showed that MaxEnt was the only consistent recipe for drawing self-consistent inferences from data. SJ only asserted, by means of four axioms, that any variational procedure must yield a unique probability distribution that satisfies the rules of addition and multiplication for independent probabilities if data does not couple the probabilities for different events.  The arguments of SJ are quite general as they do not assign explicit meaning --and, in particular, any thermodynamic meaning-- to $H$ itself. Thus $H$ can be used as a variational function to discriminate between models across a broad range of problems.  

However in recent years, other mathematical functions of $\{p_k\}$ also called entropies ~\cite{renyi, hanel, jizba, tsallis88, HottaJoichi99,TsallisRev2001, tsallis91, dosSantos, AbePhysLett} 
 --and, more broadly, regularization schemes~\cite{tiko} of which entropy maximization is a special type--
have been used to infer complex models, often power laws, from data.  
In general, these entropies violate one or more of SJ's axioms and, as a result, may be nonadditive. 
By contrast, the BG entropy is additive in the sense that
for two independent systems $A$ and $B$, the value of the BG entropy $H$ for the combined system satisfies $H(A,B)=H(A)+H(B)$.  

Nonadditive entropies have been of particular interest because they are commonly invoked when microscopic components of a system have long-range interactions.  
These unconventional entropies satisfy different properties from those of BG. As an example, according to Tsallis, Gell-Mann and Sato ~\cite{TsallisPNAS2005}, the Tsallis entropy~\cite{tsallis88, hanel, TsallisPNAS2005, TsallisJSP98, TsallisRev2001} of a scale-invariant system is extensive. 
While nonadditive entropies do not satisfy the additivity rule, they may instead satisfy a ``pseudo-additivity rule"~\cite{AbePRE}, $H(A,B)= H(A) + H(B) + \epsilon H(A)H(B)$, where $\epsilon$ is a measure of the deviation from additivity.  Nonadditive forms of the entropy function used within variational principles can preferentially select models having power law distributions~\cite{TsallisRev2001,ChoScience2002} which arise from a variety of natural and social systems.

Nonadditive entropies have been criticized on the basis, for example, that the Tsallis parameter $q$ 
\cite{wilk, beck} (related to the $\epsilon$ presented earlier) is often chosen by fitting data, rather than by some first principle~\cite{ChoScience2002}.  
Furthermore, unconventional averages must often be used to constrain nonadditive entropies ~\cite{plastino,abebagci, abeconstr, abeconstr2, abeconstr3, abeconstr4} to assure the convexity of those functions if they are to be used to infer a unique model.  

In regard to these criticisms, if the matter at hand concerns situations in which a full distribution of data is already known --and thus $q$ could be fit to data-- then it is fair to ask whether there is a need for any variational principle for selecting a model in the first place. 
This raises the question of how to justify --and when to use-- BG versus nonadditive entropies. 

The question of interest here is how nonadditive entropies can be justified within SJ's axiomatic framework. 
This framework is not about specifying properties of $H$ based on physical (and often thermodynamic) properties of an 
entropy but rather about making self-consistent inferences from data without imposing structure on a model which is not warranted by the data itself.  
Thus SJ's axioms are stronger conditions than additivity conditions on $H$.

We use SJ's reasoning to shed light on what variant of the basic logical consistency requirements are necessary to derive an alternate formula for the entropy.  
We first review SJ's axioms and show how BG's $H$ follows from the the product rule, $p_{ij} = u_i v_j$ in the absence of data coupling events $i$ and $j$.
We will then show what rules of probability would be required instead in order to justify the Tsallis entropy as well as other entropies.  
In particular, we will show from SJ that the Tsallis entropy can only be justified if events $i$ and $j$ were to have the following joint probability $p_{ij}^{q-1}
= \left(u_{i}^{q-1}+v_{j}^{q-1}-1\right)$ --presupposed in the absence of data coupling events $i$ and $j$--
rather than $p_{ij} = u_i v_j$. 

{\it The Shore and Johnson Axioms --} 
SJ considered the problem of extracting a model from data using a variational function $H(\{p_{k}\})$.   The model, $\{p_{k}^*\}$, is the one that gives the maximum of
\be
H(\{p_{k}\}) - \lambda \left( \sum_{k}p_{k}a_{k}-\bar{a}\right)
\label{maxf}
\ee
with respect to $\{p_{k}\}$ and the Lagrange multiplier $\lambda$. See Ref.\cite{abeconstr, abeconstr2, abeconstr3, abeconstr4} for a discussion of constraints.
Here the data is imposed as a constraint on the quantity $a$, where $\bar{a}$ is the measured average.
For simplicity, we considered here only a single equality constraint.  

SJ gave four axioms that must be satisfied by the maximum of the function given by Eq.~(\ref{maxf}) on the basis of requiring that any inference drawn from this function be self-consistent.  
These four axioms determine the form of $H$.

{\bf 1) Uniqueness --} says that the function $H(\{p_{k}\})$ must be convex, so that there will only be a single maximum, i.e. a single set of values, $\{p_{k}^*\}$.

{\bf 2) Coordinate system invariance --} says that predictions made from an inference should be independent of the choice of coordinate system. It is relevant when the probabilities are continuous functions and determining the dependence of $H$ on the prior over $p_{i}$.

{\bf 3) Subset independence --} says that if probability, $p_k$, of bin $k$ increases by $\delta p$ and the probability $p_j$ of bin $j$ correspondingly decreases by $\delta p$, then no other bins are affected by the change. 
Subset independence yields the relationship, 
\be
H=\sum_{k}f(p_{k})+C,
\label{subset}
\ee
where $C$ is a constant  independent of $p_{k}$.

{\bf 4) System independence --} says that bringing together two  systems having probabilities ${\bf u}=\{u_{i}\}$ and ${\bf v}=\{v_{j}\}$ gives new bins that have probability ${\bf p} ={\bf u }\times {\bf v }$ where $p_{i j}=u_{i}v_{j}$.  
The systems are considered independent if constraints on the data do not couple them.  Consider a combined system with two decoupled constraints, one on $u_{i}$ (which is $\sum_{i,j} p_{i j} a_{i}-\bar{a} = \sum_{i} u_{i} a_{i}-\bar{a} = 0 $) and another on $v_{j}$ (which is $\sum_{i,j} p_{i j}-\bar{b} = \sum_{j} v_{j} b_{j}-\bar{b} = 0$).

The maximum of 
\be
H({\bf p}) - \lambda_{a} \left(\sum_{i,j} p_{i j} a_{i}-\bar{a} \right)  - \lambda_{b} \left(\sum_{i,j} p_{i j} b_{j}-\bar{b} \right)
\label{indconstr}
\ee 
with respect to $p_{ij}=u_{i} v_{j}$ satisfies 
\be
f'(p_{ij}) - \lambda_{a}a_{i}  - \lambda_{b}b_{j} =0.
\label{next}
\ee
Taking two derivatives of the above (one with respect to $u_{i}$ and another with respect to $v_{j}$) yields
\be
f''(p_{ij}) +  p_{ij} f'''(p_{ij})=0.
\label{sjform}
\ee
We define $f''(p_{\alpha})\equiv g(p_{\alpha})$, where $\alpha \equiv (i,j)$.  Then Eq.~(\ref{sjform}) reduces to $g(p_{\alpha})+p_{\alpha}g'(p_{\alpha})=0$.
The solution is $g(p_{\alpha})=-1/p_{\alpha}$.  It follows that $f(p_{\alpha}) = -p_{\alpha}\log p_{\alpha} +p_{\alpha}$ and $H=-\sum_{\alpha} p_{\alpha}\log p_{\alpha} +C$, where all additional constants have been absorbed into $C$. 

The derivation above shows how the BG formula follows from the axioms of SJ. 
Intuitively, SJ's axioms {\bf 3} and {\bf 4} take as definitions the fact that events are independent unless the data couples them and the probabilities for independent events satisfy the usual rules of addition and multiplication for such probabilities.  
This explains why the BG entropy is additive.

However, not all physical systems are additive; often cited as counterexamples are systems having long-ranged interactions~\cite{TsallisRev2001}. The question is how should nonadditivity be built into a model?
One route has been to redefine entropy and replace it with a form which violates axiom {\bf 4}, system independence -- the law of multiplication of probability for independent events, $p_{k}= u_{i} v_{j}$. 
Here we demonstrate the logical consequences that follow from redefining the entropy.

The Tsallis entropy is defined as
\be
H=\frac{K}{1-q} \left(\sum_{k}p_{k}^{q} -1\right).
\label{tsallisent}
\ee
This expression satisfies subset independence, axiom {\bf 3}, but does not satisfy system independence, axiom {\bf 4}. What functional form for $p_{ij}= p (u_{i}, v_{j})$ yields the Tsallis entropy?
To answer this question, we repeat steps analogous to those in Eqns.~(\ref{indconstr})-(\ref{sjform}) except now we treat $p_{ij}$ as a general function of $u_{i}$ and $v_{j}$ and
$f(p_{ij})$ is given by the form Eq.~(\ref{tsallisent}). This gives 
\be
(2-q)^{-1}  p_{ij}\frac{\partial^{2} p_{ij}}{\partial u_{i}\partial v_{j}}  = \frac{\partial p_{ij}}{\partial u_{i}} \frac{\partial p_{ij}}{\partial v_{j}}.
\label{diffp}
\ee
Eq.~(\ref{diffp}) is a differential equation satisfied by the joint probability in the Tsallis entropy.  As a check, we can see that for $q=1$ --when the Tsallis entropy reduces to the BG entropy-- 
the expression is exactly satisfied for $p_{ij}=u_{i}v_{j}$, as expected. The constant $(2-q)^{-1}$ in Eq.~(\ref{diffp}) describes the deviation from independence (often one speaks of deviation from independence in terms of the q-additivity of the Tsallis entropy --as opposed to normal additivity of entropy in statistical mechanics~\cite{tsallis91, AbePRE, AbePhysLett}.). 

We now solve Eq.~(\ref{diffp}).\\
{\bf Step 1 --} 
Substitute $p_{ij} = h_{ij}^{x}$ --where $x$ is a number-- into Eq.~(\ref{diffp}). 
After some algebraic rearrangement, this yields
\be
-(2-q)^{-1} x h_{ij} \frac{\partial^{2} h_{ij}}{\partial u_{i}\partial v_{j}}
=\left(x(x-1)(2-q)^{-1}-x^{2}\right) 
\frac{\partial h_{ij}}{\partial u_{i}}\frac{\partial h_{ij}}{\partial v_{j}} 
\label{step1}
\ee
We select $x$ such that $\left(x(x-1)(2-q)^{-1}-x^{2}\right) =0$.
The non-trivial solution to this quadratic equation is $x=1/(q-1)$
(the trivial solution is $x=0$). Plugging $x=1/(q-1)$ into Eq.~(\ref{step1}),
 we have
\be
\frac{\partial^{2} h_{ij}}{\partial u_{i}\partial v_{j}} = 0.
\label{step1b}
\ee
Eq.~(\ref{step1b}) is solved by
$h_{ij} = \phi_{1}(u_{i})+\phi_{2}(v_{j})$ with $\phi_{1}$ and $\phi_{2}$ yet to be determined.
Since both $p_{ij}$ and $h_{ij}$ must be symmetric functions of $u_{i}$ and $v_{j}$,
then $\phi_{1}=\phi_{2}\equiv \phi$.
We therefore have
\be  
p_{ij}^{q-1}=h_{ij} = \phi(u_{i})+\phi(v_{j})
\label{step1c}
\ee
{\bf Step 2 --} In order to determine the function $\phi(x)$, we rewrite it is as
\be
\phi(x) = g(x)^{q-1} - 1/2 
\ee
without loss of generality, so that Eq(\ref{step1c}) takes the form
\be  
p_{ij}=\left( g(u_{i})^{q-1}+g(v_{j})^{q-1} -1 \right)^{1/{(q-1)}}.
\label{step1d}
\ee
The leading order expansion of (\ref{step1d}) in $q-1$ is 
\be
p_{ij} = g(u_{i})g(v_{j}) - (q-1)g(u_{i})g(v_{j}) \log g(u_{i})\log g(v_{j}) + O((q-1)^2).
\label{expl2}
\ee
and from the condition that the composition rule reduces to the product rule in the limit of $q \to 1$, we get $g(x)= x$. 
Substituting it back into Eq.(\ref{step1d}), we get 
\be
p_{ij}=\left(u_{i}^{q-1}+v_{j}^{q-1}-1\right)^{1/(q-1)}.
\label{pkq}
\ee
Eq.(\ref{pkq}) can also be written in terms of the q-product defined by Tsallis\cite{TsallisRev2001}.
Furthermore,  Eq.~(\ref{diffp}) gives us the choice to select a multiplicative constant that can be 
determined by normalization of the joint probability, $p_{ij}$. 
Fundamentally, the spurious correlations between events --which consist of all terms beyond the first term in Eq.~\ref{expl2}-- 
emerge because the Tsallis entropy violates SJ's axiom {\bf 4}.
This axiom specifically requires that in the absence of couplings between events $i$ and $j$, the model inferred using the BG entropy
should satisfy the normal rules of multiplication of probability ($p_{ij}=u_{i}v_{j}$). 

Many entropies --beyond the Tsallis entropy-- also violate axiom {\bf 4}. These entropies therefore generate spurious correlations not warranted by the data
even if they are additive in the sense $H(\{p_{ij}\}) = H(\{u_{i}\})+H(\{v_{j}\})$. 
In other words, axiom $4$ is a stronger statement than is the statement that $H$'s add.

For instance, consider the Burg entropy\cite{karlin}  $K\sum_{k}\log p_{k}$ which is additive in the sense described above.
The Burg entropy satisfies SJ's axiom {\bf 3} but violates axiom {\bf 4}. 
For this entropy, the system dependence relationship still deviates from the rule of multiplication of probability for independent events
\be
p_{ij}^{-1}=u_{i}^{-1}+v_{j}^{-1}-1.
\label{others}
\ee

Eqns.~(\ref{pkq}) and ~(\ref{others}) underscore the profound consequences that result from altering the form of the entropy used in model inference.
SJ's axioms assure us that the BG form of the entropy enforces a model distribution which is as featureless as possible. 
According to SJ's framework, couplings between events --or more broadly, structure in a model-- arises in one of two ways. 
Either the couplings in the data explicitly give rise to correlations between events $i$ and $j$ or the prior over the $\{p_{k} \}$ --the set $\{ q_{k} \}$
which can be thought of as a hyperprior-- gives rise to structure beyond what is present in the data. 
Thus application of the BG entropy ensures that inferences do not go beyond what is in the data or $\{ q_{k} \}$. 

However non-traditional entropies, which violate SJ's axiom ${\bf 4}$,  are inconsistent with the probability relationship $p_{ij} = u_i v_j$ 
even in the absence of any evidence of coupling between events $i$ and $j$.
While entropy priors --such as the Tsallis entropy-- can readily infer power law distributions for $\{p_{k} \}$,
they impose structure in a model that goes beyond what is known from the data. Here Eqns.~(\ref{pkq}) and ~(\ref{others}) 
derive this additional structure imposed by these entropies on a model explicitly.
We conclude by adding that it is possible to infer power laws within a principle of maximizing the BG entropy by constraining just one average: 
Mandelbrot~\cite{mandelbrot} showed this by invoking logarithmic constraints, $\langle \log k \rangle$.  

In summary, the maximization of entropy is a variational prescription for selecting one of many possible models of probability distributions consistent with limited data.  In a seminal result that we review here, SJ showed that only the BG entropy or functions with identical maxima ensure that models derived from them satisfy basic logical self-consistency requirements.  We apply SJ's approach to derive what joint probability for states of two systems would be required to justify the form of the Tsallis entropy as well as other entropies in selecting model probability distributions consistent with data.  We observe that all forms of nonadditive entropy functions require probability rules other than the multiplication rule even when events are independent according to data. We conclude that for modeling nonexponential distributions, such as power laws, nonextensivity should be expressed through the constraints or the $q$'s, not the entropy.  Said differently, no structure should be assumed in a distribution function unless it is observed as coupling in the data or originates from the prior distribution on $\{p_{k}\}$.

\section{Acknowledgements}
We thank the referees for their insightful feedback.
SP acknowledges support from the Purdue Research Foundation as well as support from his IUPUI Start-up.
KD acknowledges support of NIH grant 5R01GM090205-02 and the Laufer Center. 
K.G. acknowledges support from the Research Corporation for Science Advancement (as a Cottrell Scholar), National Science Foundation (Grant No. 1149992) and PROF grant from the University of Denver.

\bibliographystyle{unsrt}

\end{document}